\documentclass[conference,10pt]{IEEEtran}
\usepackage{amsmath}
\usepackage{amssymb}
\usepackage{amsfonts}
\usepackage{graphicx}
\usepackage{enumerate}
\usepackage{bbm}

\usepackage{mathabx} 
\usepackage{flushend}
\usepackage{cite}
\usepackage{subfigure}
\usepackage{url}
\usepackage{array}
\usepackage{verbatim}
\usepackage{comment}
\usepackage{stfloats}

\usepackage{color}
\definecolor{red}{rgb}{1,0,0}
\definecolor{blue}{rgb}{0,0,1}

\long\def\comment#1{}

\newfont{\bbb}{msbm10 scaled 700}

\newfont{\bb}{msbm10 scaled 1100}

\newcommand{\detus}[2]{D_{#1,#2}}

\newcommand{\detsec}[1]{{\cal D}_{#1}}
\newcommand{\uicode}[1]{{\cal K}_{#1}}

\newcommand{\presus}[2]{X_{#1,#2}}

\newcommand{\presec}[1]{{\cal X}_{#1}}

\begin{document}
\title{
A novel alternative to Cloud RAN for throughput densification: Coded pilots and fast user-packet scheduling at remote radio heads}

\author{\IEEEauthorblockN{Ozgun Y. Bursalioglu$^1$, Chenwei Wang$^1$, Haralabos Papadopoulos$^1$, \\Giuseppe Caire$^2$}
\IEEEauthorblockA{$\ ^1$Docomo Innovations Inc, Palo Alto, CA\\
$\ ^2$Technische Universit{\"a}t Berlin, Germany\\
\{obursalioglu, cwang, hpapadopoulos\}@docomoinnovations.com,\, caire@tu-berlin.de}}

\maketitle
\begin{abstract}
We consider wireless networks of remote radio heads (RRH) with large antenna-arrays, operated in TDD, with uplink (UL) training and channel-reciprocity based downlink (DL) transmission. To achieve large area spectral efficiencies, we advocate the use of methods that rely on rudimentary scheduling, decentralized operation at each RRH and user-centric DL transmission.

A slotted system is assumed,  whereby users are randomly scheduled (e.g., via shuffled round robin) in slots and across the limited pilot dimensions per slot. As a result, multiple users in the vicinity of an RRH can simultaneously transmit pilots on the same pilot dimension (and thus interfering with one another). Each RRH performs rudimentary processing of the pilot observations in ``sectors''. In a sector, the RRH is able to resolve a scheduled user's channel when that user is determined to be the only one among the scheduled users (on the same pilot dimension) with significant received power in the sector. Subsequently, only the subset of scheduled users whose channels are resolved in at least one sector can be served by the system.

We consider a spatially consistent evaluation of the area multiplexing gains by means of a Poisson Point Process (PPP) problem formulation where RRHs, blockers, scatterers and scheduled user terminals are all PPPs with individual intensities. Also, we study directional training at the user terminals. Our simulations suggest that, by controlling the intensity of the scheduled user PPP and the user-pilot beam-width, many fold improvements can be expected in area multiplexing gains with respect to conventional spatial pilot reuse systems.
\end{abstract}



\vspace{-0.2cm}
\section{Introduction}
\label{section:introduction}
\vspace{-0.15cm}

A broad range of activities are currently underway towards realizing the ever-evolving vision of 5G. Their scope includes the development and standardization of new technologies in 3GPP  \cite{3GPP}, new channel models \cite{5GSCM}, experimental trials and demos,  and the study of new services and their requirements \cite{NGMN}.  5G technologies are expected to bring great performance gains with respect to their predecessors in a multitude of performance metrics, including, among other things, user and cell throughput, end-to-end latency, massive device connectivity and localization. They are also viewed as essential elements for enabling the much broader gamut of services envisioned, such as immersive applications (e.g., virtual/augmented/mixed reality) \cite{Qi_Immersive, Medard_5GVR}, haptics \cite{Dohler_Haptic}, V2X \cite{V2X} and the Internet of Things \cite{5GIoT}. To meet such an ambitious and broad gamut of 5G requirements, operators would have to rely on a combination of additional resources, which include newly available licensed and unlicensed bands, network densification, large antenna arrays and new PHY/network layer technologies. Also, to meet the increased traffic demands per unit area induced by the multitude of new services and their requirements, 5G systems would need to provide much higher area spectral efficiencies and area multiplexing gains (e.g., number of users streams served simultaneously per unit area) than their 4G counterparts.

Antenna densification through massive MIMO is expected to play a key role in achieving the large gains in area spectral efficiency envisioned for 5G and beyond wireless networks. Through the use of a large number of antennas at the base stations (BSs), massive MIMO can yield both large spectral efficiencies and spatial multiplexing gains \cite{marzetta2006, marzetta-massive}.
Large antenna arrays and massive MIMO are considered as key technologies for 5G and beyond, in particular because new generation deployments would have to utilize the centimeter and millimeter Wave (mmWave) bands where wide chunks of spectrum are readily available.
Indeed, at mmWave with half-wavelength (critical) spacing, very large arrays can be packed on small footprints. Realizing the beamforming gains that can be provided by such large-size arrays will be essential in combatting the harsher propagation characteristics experienced at mmWave.

To achieve large spectral efficiencies in the downlink (DL) via multiuser (MU) MIMO, channel state information at the transmitter (CSIT) is needed \cite{caire2010multiuser}. Following  \cite{marzetta-massive}, CSIT can be obtained from the users' uplink (UL) pilots via Time-Division Duplexing (TDD) and UL/DL radio-channel reciprocity. This approach allows training large arrays by allocating as few UL pilot dimensions as the number of single-antenna users simultaneously served.  

At higher carrier frequencies,  UL-pilot based  CSI acquisition becomes even more attractive than schemes relying on DL training and UL CSI feedback.  First,  at higher frequencies larger BS arrays can be packed within a fixed footprint, and, since a single UL pilot from a  user terminal trains all the BS antennas (no matter how many), UL-pilot  based schemes are able to harvest the additional BF gains available at higher frequencies without additional overheads \cite{marzetta2006}.  Second,  UL-pilot based schemes also induce lower latencies in  CSI acquisition than feedback based schemes \cite{marzetta2006}. This is a key advantage at higher frequencies where the latency requirements in CSI acquisition are more stringent, as CSI is only accurate within the coherence time of the channel, and coherence time is inversely proportional to carrier frequency.

In this paper, we focus on operating a wireless network of remote radio heads (RRHs) with large antenna-arrays.  We consider TDD operation and DL data transmission, enabled by UL training and UL/DL radio channel reciprocity \cite{scalable-sync-cal-2014a}. To achieve large area multiplexing gains and area spectral efficiencies, we investigate the combined use of rudimentary scheduling, decentralized operation at each RRH and user-centric DL transmission. Specifically, we consider a slotted system where the network schedules a fraction of users on each of the UL pilot dimensions in each slot. As a result, multiple users in the vicinity of an RRH can transmit UL pilots on the same pilot dimension and can cause pilot contamination. With pilot-contaminated CSI, a beam directed to an intended user terminal  also inherently inherently beamforms at other  user terminals using the same pilot dimension,  causing interference and substantially limiting massive MIMO performance.

To address the issue of pilot contamination, we consider rudimentary processing of the pilot observations at  each RRH, which allows each RRH to resolve user channels in ``sectors.''  An RRH is able to resolve an active (scheduled) user's channel in a sector when that user is determined as the only one among the active users (on the same pilot dimension  with significant received power) within that sector. Subsequently, only the subset of active users whose channels are resolved in at least one sector can be served. Regarding the type of sectorization, we study physical sectorization, virtual sectorization and their combined use. While physical sectorization refers to the conventional sectorization where each sector is operated by a distinct antenna array, virtual sectorization arises through sector-specific spatial filtering from a single antenna-array. 

In order for each RRH to resolve user channels, special pilot coding is needed for active users who send pilots over the common pilot dimension, i.e., across the set of shared resource elements (REs) for UL pilots.
Recently, Bursalioglu {\it et al.} proposed ON-OFF type of non-orthogonal pilot codes, which use a small fraction of additional REs per pilot dimension to detect pilot collisions and to identify the user IDs whose channels can be resolved on the fly \cite{icc2016-ozgun}.  Other classes of non-orthogonal pilot codes were described by Li {\it et al.} \cite{icc2017-zheda}, which assume wide-band scheduling transmission 
and are able to resolve user channels using codes without incurring additional overheads.

In this work, we conduct a spatially consistent system-level evaluation of the area multiplexing gains by these systems. Besides sectorization at RRHs, we also consider the use of {\em directional training at the user terminals} and its effect on the provided area multiplexing gains. Given an appropriately chosen beam-width (common across all user terminals), each active user picks a random beam-orientation to transmit its pilot on its assigned pilot dimension. Similar type of directional training was exploited in the context of cellular transmission in \cite{icc2017-zheda}. By taking advantage of the sparsity of mmWave channels in the angular domain, significant gains were reported in terms of both area multiplexing gains and area spectral efficiencies. 

 We consider a PPP-based problem formulation where RRHs, blockers, scatterers and user terminals are all PPPs with individual intensities. We also assume that the scheduled user terminals form a PPP with an intensity that can be adjusted to optimize the area multiplexing gains. Indeed, random user scheduling in this context corresponds to appropriate thinning of the original user terminal PPP.
 
We conduct extensive simulations by varying the type and extent of RRH sectorization, the user pilot beam-width and the user scheduling intensity. By adjusting the intensity of the scheduled-user PPP and the user-pilot beam-width, significant improvements in area multiplexing gains are reported with respect to conventional spatial pilot reuse systems. These gains greatly improve with RRH sectorization and are present if one employs physical sectorization, virtual sectorization or a combination.

The PPP-based ON-OFF connectivity model we employ in this work can be viewed as a realistic extension of the model in \cite{icc2016-ozgun}. The model relies on a simple connectivity model involving LOS and NLOS paths. A link is ``ON'' if that link strength exceeds a predetermined threshold. Thus, the LOS link between a user and an RRH is ``ON'' if the line of sight path is not blocked and the LOS received path strength exceeds a predetermined threshold. Similarly,  NLOS paths are ``single-bounce'' through a single scatterer with sufficiently high received path strength. We remark that, although our model is very simple, it captures the main aspects of the problem and provides spatial consistency of the type not immediately present in stochastic channel models such as 3GPP's SCM \cite{3DSCM}. In addition, by controlling the intensities of the underlying PPPs, important statistics of our model (such as, e.g., number of paths between a connected user-BS link) can be tuned to agree with those in the stochastic 3GPP SCM model.

\section{Channel model}
\label{channel-model}

In this section, we provide a spatially consistent ON-OFF RRH connectivity model capturing signal attenuation for both LOS and NLOS paths.
In particular, we consider a distance-based\footnote{The distance metric we consider in this paper is defined as the Euclidean distance in the two-dimensional (X,Y)-plane. The vertical dimension is not taken into account.} signal attenuation model for a direct LOS path between a user $u$  and an RRH $r$. Let $d_{ur}\ge 0$ denote the distance between user $u$ and  RRH $r$, and let
\[
b_{ur} = \begin{cases} 0 & \text{if the LOS path between  $u$ and $r$ is blocked},\\
1 & \text{otherwise.}
\end{cases}
\]
Then the path attenuation on the direct path between user $u$ and RRH  $r$ is modeled via $g(d_{ur})\, b_{ur}$,. We choose a smooth transition path loss function $g(\cdot)$ \cite{antenna_book}  given by
\begin{equation}\label{LOS}
g(d) = \left(1+d/\epsilon\right)^{-\alpha}, \ \ \ \ \  \text{for $d\ge 0$}, 
\end{equation}
where $d$ denotes the distance between transmitter and receiver,  and $\epsilon$ and $\alpha$ denote the breakpoint distance and the pathloss exponent, respectively\footnote{We remark that  $g(d_{ur})$ refers to the 	``directional'' propagation loss, i.e., attenuation that is specific to the path between user $u$ and RRH $r$. This is different from the commonly used distance-based pathloss models such as, e.g., 3GPP's SCM \cite{3DSCM},  which are modeling the omni pathloss, i.e., the  aggregate propagation loss across {\em all} paths.}.

Simple inspection on (\ref{LOS}) reveals that $g(d)$ possesses the following desirable properties:  (i) it is a monotonically decreasing function of the distance $d$, implying longer propagation paths are more attenuated; (ii) it satisfies  $0\le g(d)\le g(0)=1$ showing that propagation always attenuates signal strength.

Regarding NLOS paths, we restrict our attention to single-bounce paths through a single scatterer. We consider a commonly used method for modeling reflected path attenuation \cite[Chapter 2]{wc_book_tse}, according to which the total attenuation of a reflected path is given as the product of the attenuation levels experienced by the individual paths.  In particular,  the path attenuation in a NLOS path between user $u$ and  RRH $r$ enabled through a single-bounce reflection off a scatterer $z$ is given by
\begin{equation}\label{NLOS}
f(d_{uz},d_{zr}) = a\,g(d_{uz}) \, g(d_{zr}),
\end{equation}
where $a\le 1$ is the reflector-attenuation factor,  and $d_{uz}$ and $d_{zr}$ denote the user-scatterer and the scatterer-RRH distances, respectively.
It can be readily  verified that the single-bounce path attenuation model (\ref{NLOS}) with $g(\cdot)$ defined in (\ref{LOS}) has the following desirable properties:
\begin{itemize}
\item [(i)]  For any single-bounce path,  $0\le f(d_{uz},d_{zr}) \le 1$.  In addition, $f(d_{uz},d_{zr}) \to 1$, if and only if $d_{uz},\,d_{zr}\to 0$ and $a \to 1$.
\item [(ii)]  The attenuation scaling of a single-bounce path is a non-increasing function of the user-scatterer distance and of the scatterer-RRH distance: $f(d_{uz},d_{zr}) \ge f(d'_{uz},d'_{zr})$, for any  set of  distances \{$d_{uz}$,  $d'_{uz}$, $d_{zr}$, $d'_{zr}$\} satisfying $0\le d_{uz} \le d'_{uz}$ and $0\le d_{zr}\le d'_{zr}$.
\item[(iii)] Considering the triangle formed by the LOS paths connecting the points $u$, $z$, and $r$, we have  $d_{uz}+d_{zr}\geq d_{ur}$ and $f(d_{uz},d_{zr})\leq g(d_{ur})$ as well. Hence, when not blocked, the received signal strength of the direct LOS path is at least as large as any reflected path, even when $a = 1$, i.e.,  even when there is no reflector-attenuation. 
\end{itemize}

Coverage is defined by means  of a simple ON-OFF model: user $u$ is within the coverage of RRH $r$, if and only if there exists a path, either a LOS path or a NLOS path, received with sufficiently large strength. Specifically, user $u$ is within the coverage of RRH $r$, if $g(d_{ur})\, b_{ur}\ge \delta$ or if there exists a scatterer $z$ for which $f(d_{uz},d_{zr})\geq \delta$,  for some  predetermined threshold $\delta$. In contrast, a user is in outage if it is not in the coverage of any RRH in the system.

It is worth denoting by $d_o$ the nominal distance at which $g\left(d_o\right)=\delta$, which implies that if $b_{ur}=1$ a user $u$ at distance $d_{ur}<d_o$ from RRH $r$ is in coverage of the RRH. In contrast, if user $u$ is at distance $d_{ur}> d_o$ from RRH $r$, the user is not in coverage of the RRH. Hence, the disk centered at user $u$ with radius $d_o$ defines the support of all RRHs with respect to which user $u$ {\em could} be in coverage. Similarly, from the RRH side, all the user terminals that could potentially be in the coverage of RRH $r$ have to reside in the radius-$d_o$ disk centered at RRH $r$.

\section{System Operation}

We consider a slotted system where a subset of users are scheduled in each slot for UL pilot training and subsequent DL data transmission. We let $\tau$ denote the number of available orthogonal pilot dimensions per fading block, and assume each active user is scheduled on one of the $\tau$ pilot dimensions.

We will consider the use of both omni and directional UL pilots by active users. In the omni-case, we assume  that a user's UL pilot coverage coincides with the user coverage in Sec.~\ref{channel-model}. Directional pilot transmissions are parameterized by a positive integer $B\ge 1$, which remains common and fixed for all user terminals. Given a $B\ge 1$, the user terminal creates $B$ non-overlapping pilot beams that span the azimuth spectrum (360 degrees). In the simplest ``geometric" case, each beam has beam-width $360/B$ degrees. During each scheduling instance, each active user transmits an UL pilot on randomly selected one of the $B$ beams on the assigned pilot dimension. 

To ensure that coverage is identical for both omni and directional pilot training, the UL pilot of each directional beam is adjusted, so that the union of all $B$-beam coverage areas coincides with the omni-pilot coverage, i.e., a disc radius of $d_o$.  This approach ensures a fair comparison, and thus any performance gains arising from the use of directional pilot training cannot be attributed to coverage extension.

Upon transmission of UL pilots, per-sector processing at each RRH  allows the RRH to resolve user channels on each of its sectors and subsequently serve the resolved user streams. A user channel is considered resolved on a sector, if the user channel is present on this sector (i.e. the user has a path with sufficiently large strength on that sector) and if it is the only user present on the sector among the active ones on the same pilot dimension. In the following subsections, we describe sectorization and the notion of presence of user channels on sectors, user scheduling and resolving user channels. We also provide expressions for the achievable area multiplexing gains per pilot dimension by using these schemes.

\subsection{Sectorization}

\begin{figure}
\centering
\centerline{\includegraphics[angle=90,width=7cm]{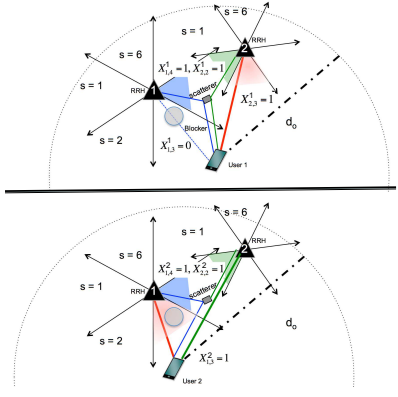}}
\vspace*{-.1in}
\caption{Two RRHs are shown within $d_o$ distance from the two UTs. The first and the second RRHs are, respectively, represented by the triangles 1 and 2. Each RRH employs 6 sectors. Also shown are a single scatterer and a single blocker.}
\label{fig_sectors}
\vspace*{-.1in}
\end{figure}

Sectorization has been traditionally employed for increasing the spectral efficiency per site in cellular networks by partitioning each cell radially into sectors and reusing the spectral resources in each sector\cite{valenzuela-sector}. We let $S$ denote the number of radial sectors per site. Fig.~\ref{fig_sectors} shows an example of uniformly spaced radial sectorization with $S = 6$ sectors. In this case, the circular coverage area centered at an RRH is divided into equal angle wedges, one per sector. We note that, in general, the $S$ sectors can be constructed by means of traditional physical sectorization, virtual sectorization, or a combination of both techniques. The impact of these options on the sectorization pattern and consequently on the system performance is considered in Sec.~\ref{antenna-array}.

As in \cite{icc2017-zheda}, we let $\presus{j}{s}^k=1$ denote that a user $k$ is present on sector $s$ of RRH $j$ and set $\presus{j}{s}^k=0$ otherwise. For omni-pilot transmission, $\presus{j}{s}^k=1$, if there exists a path from user $k$ to RRH $j$ with attenuation scaling exceeding $\delta=g(d_o)$ and with angle of arrival (AoA) falling within the support of sector $s$. For directional training to have  $\presus{j}{s}^k=1$, we require, besides the above two conditions, that the path from user $k$ to RRH $j$  have  angle of departure (AoD) falling within the support of user's UL pilot beam. Evidently, directional pilot transmissions have lower incidence of user-channel presence than omni pilot transmissions, effectively sparsifying the observed user channels at each RRH.


Fig. \ref{fig_sectors} provides an illustrative example  involving  two users located close to each other, two nearby RRHs each with geometric $S=6$ radial sectorization, a single nearby scatterer and a single nearby blocker.\footnote{We assume ideal sectorization, i.e., there is no leakage between sectors. Furthermore, all scatterers are assumed to be reflecting isotropically.}
Both users are  assumed to use omni-directional pilots and to be within $d_o$ distance of the two RRHs. In addition, all four reflected paths through the scatterer, corresponding to all (user, RRH) permutations, are also within coverage. The upper and lower figures separately show the presence of user-1 and user-2 on RRH sectors. As the top figure reveals, user 1 is present on sector 4 of RRH 1 (reflected path), and sectors 2 and 3 of RRH 2 (through reflected and direct paths, respectively). Note that user 1 is not present in sector 3 of RRH 1, as the the blocker blocks  the LOS path. Similarly, user 2 is present on sectors 3 and 4 of RRH 1 and on sector 2 of RRH 2. 

\subsection{Scheduling \& Resolving Users}

Next, we describe scheduling and resolving user channels. Since we assume wide-band scheduling, a scheduling slot comprises a set of concurrent fading blocks. Each fading block can be  viewed as spanning a contiguous set of time-frequency elements in the OFDM plane within the coherence bandwidth and time of the user channels \cite{icc2017-zheda}.

We assume $L = \tau \lambda_U A$ users are randomly scheduled for uplink pilot transmission over an area of size $A$, where $\lambda_U$ is an appropriately chosen density of scheduled users so that $L$ is an integer. For comparison, in the context of the traditional orthogonal training scheme, in each slot $L = \tau$ users are scheduled in the entire area, i.e., a single scheduled user per pilot dimension resulting in $ \lambda_U=1/A$. Thus, $\tau$ users send orthogonal pilots on each fading block, i.e., $K=1$  user per pilot dimension. On the other hand, in the context of the non-orthogonal UL training schemes, as in \cite{icc2016-ozgun, icc2017-zheda}, $K>1$ users are scheduled to send pilots per pilot dimension in the slots. As a result, the number of scheduled users per slot is $L=K\tau$.

Based on UL training, each sector of each RRH {\em resolves} the channels of a subset of the $L$ users and serves them simultaneously. Among scheduled users, only the ones whose uplink pilots are received without any ``collisions" from other user pilots can be served. Also, these are the users whose channels can be ``resolved" to be used in designing precoders. Thus, we consider a user $k$ to be {\em resolved} on sector $s$ of RRH $j$ if and only if $\presus{j}{s}^k=1$  and there is {\em no other} user $k'$  sharing the same dimension as user $k$ and for which $\presus{j}{s}^{k'}=1$.
Letting $\sigma_k$ denote the pilot dimension used by user $k$, $\uicode{\sigma}$  denote the indices of users assigned to pilot dimension $\sigma$ for $1\le \sigma\le  \tau$, and
$\detus{j}{s}^k$ denote whether or not user $k$'s channel can  be {\em resolved} on sector $s$ of RRH $j$, we have \cite{icc2017-zheda}:
\begin{equation}
\label{detus-def}
 \detus{j}{s}^{k} = \presus{j}{s}^{k}\left[ \prod_{k'\in \uicode{\sigma_k}\setminus \{k\}} \left(1-\presus{j}{s}^{k'}\right) \right].
\end{equation}
The subset of present users whose channels are resolvable in sector $s$ of RRH $j$ is thus given by
 \begin{equation}
\label{detsec-def}
\detsec{j,s}  = \left\{ k; \ \detus{j}{s}^k = 1\right\}\ .
\end{equation}
Assuming the two users in Fig. \ref{fig_sectors} are scheduled on the same pilot dimension, it is evident that $ \detus{2}{3}^{1} = 1$ and $ \detus{1}{3}^{2} = 1$. This toy example demonstrates the effect of sectorization in increasing area multiplexing gains by being able to simultaneously serve two nearby users even if they are using the same pilot dimension.

In general, not all present users are resolvable. Letting
\begin{equation}
\label{presec-def}
\presec{j,s}  = \left\{ k; \ \presus{j}{s}^k = 1\right\}
\end{equation}
denote the set of all users that are present in sector $s$ of RRH $j$, we have $\detsec{j,s} \subseteq \presec{j,s}$. Indeed, inspection of (\ref{detus-def}) reveals that if the two users $k$ and $k'$ are present in sector $s$ and use the same pilot dimension, i.e., $\presus{j}{s}^k  = \presus{j}{s}^{k'}=1$ and $\sigma_k  =\sigma_{k'}$, we have $\detus{j}{s}^k  = \detus{j}{s}^{k'}=0$. This is consistent with the fact that neither channel can be resolved  due to the pilot collision. Hence, the number of sectors that can resolve and thus serve user $k$ is given by $N_k = \sum_{j}\sum_{s=1}^S \detus{j}{s}^k$, while the number of users that are actually served in the slot is given by
\begin{equation}
\label{MGinst-def}
L' = \left| \{ k;   N_k >0 \} \right|
\end{equation}
and in general $L'\le L$.

%

We define the slot-averaged area multiplexing gain per pilot dimension (which is a function of scheduled user density $\lambda_U$ and area size $A$) as follows:
\begin{equation}
\label{MGinst-sample-avg}
{\rm MG(A,\lambda_U)}=\lim_{T\to \infty} \frac{1}{T} \sum_{t=1}^T \frac{L'(t)}{\tau A},
\end{equation} where $L'(t)$ represents the instantaneous multiplexing gain over slot $t$ in the area $A$ and is given by (\ref{MGinst-def}).

With a traditional orthogonal scheme, since $L'(t)= \tau$ we have ${\rm MG(A,1/A)}=1/A$, which becomes vanishingly small as $A\to \infty$. This is well recognized in traditional networks and is dealt with conventional spatial pilot reuse. Such spatial reuse scheme would be inherently limited  by the nominal user-pilot coverage area $\pi d_o^2$. Hence, the orthogonal scheme with coverage area $\pi d_o^2$  provides {\em an upper bound} on the area multiplexing gains  per pilot dimension provided by {\em any} such conventional cellular deployment with conventional spatial reuse, and which is equal to $1/(\pi d_o^2)$ user streams per unit area and per pilot dimension.


\section{PPP-based Analysis}

In this section, we analyze the proposed schemes in the limit where the network size $A$ goes to infinity. We exploit a  PPP-based formulation  to  analyze the area multiplexing gains per pilot dimension provided by the proposed schemes.

In a PPP-based layout, points are assumed to be dropped over an infinite area according to a PPP with a certain intensity per unit area. We choose the unit area as $\pi d_o^2$, which is equal to the nominal coverage area that can be obtained by an omni-directional pilot from a user. Users, scheduled users, RRHs, scatterers and blockers are dropped according to PPPs with intensities $\lambda_U^{\rm all},\lambda_U,\lambda_R, \lambda_S, \lambda_B$, respectively. The scheduled user PPP with intensity $\lambda_U$, arises as thinned version of the PPP corresponding to all of the users in the network, with intensity $\lambda_U^{\rm all}$. While RRHs, users and scatterers are considered as points on the layout, the blockers are considered as circles with an area size of $1/A_b$. In the extreme case $\lambda_S = \lambda_B = 0$, i.e., when there are no scatterers or blockers, the PPP setup reduces to the one used in  \cite{icc2016-ozgun}, \cite{Haenggi}  where area multiplexing gain expressions were derived via a stochastic geometry based analysis.  Unlike  \cite{Haenggi} where exact closed-form expressions are derived for a 1-dimensional line network,  in this paper we consider the 2-dimensional case. 
For the setting we consider, closed-form analysis based on stochastic geometry is much more complicated due to presence of sectorization, directional training, scatterers and blockers. As a result, we rely on simulation based investigations. In our simulations, we run many frames, where, in each frame, users, scatterers, RHHs and blockers are dropped according to their PPPs.

In the PPP evaluation, we are interested in the area multiplexing gains per unit area and per pilot dimension:
\begin{equation}
\label{MGinst-sample-avg}
{\rm MG(\lambda_U)}=\lim_{A\to\infty} {\rm MG(A,\lambda_U)}.
\end{equation}
Note that ${\rm MG(\lambda_U)}$ can be optimized by varying the scheduled user intensity:
\begin{equation}
\label{mq-opt}
{\rm MG^*}=\max_{0\leq \lambda_U\leq \lambda_U^{\rm all}} {\rm MG(\lambda_U)},
\end{equation}
where $\lambda_U^*$ denote the corresponding optimal value.


It is worth re-iterating that the traditional orthogonal scheme has zero area multiplexing gain. Moreover, the area multiplexing gains per pilot dimension of locally orthogonal schemes that rely on conventional reuse without any channel resolution capability are nominally upper bounded by 1 (since $\pi d_o^2=1$).

\subsection{Simulation Parameters}
\label{outage}

All the simulations of this paper are based on the following parameter values: $\lambda_R =  5$ (RRH intensity), $\lambda_B = 3$ (blocker intensity), $A_b = 20$ (inverse of blocker area), $\alpha = 4$ (pathloss exponent), $\epsilon = d_o/4$ (cutoff parameter for attenuation model), $a = 1$ (scatterer attenuation) and $\lambda_U^{\rm all} = 100$ (user intensity).

We first investigate the probability of outage. Specifically, a user $k$ is in outage if (with omni pilots) for all $s, j$, $\presus{j}{s}^{k} = 0$.\footnote{With directional training, in every scheduling slot. a user selects a beam at random from a set of beams whose union covers the whole 360 degree azimuth domain. If a user is not in outage with omni-training, with directional training it will be present at one sector with at least one of its directional beams. Thus, it will not be in outage with directional training as well.}
The outage probability depends on $\lambda_R, \lambda_S, \lambda_B$ and $A_b$:  Larger $\lambda_R$ or larger $\lambda_S$ values decrease the outage probability while increasing $\lambda_B$ or $A_b$ increases the outage probability. We investigate the outage probability for two different scattering environments: $\lambda_S = 50$ (low scatterer intensity) and   $\lambda_S  = 200$ (high scatterer intensity). For the
low scatterer intensity case the outage probability is $1.9 \%$, while for the higher scatterer-intensity case it is $1.3\%$.

\begin{figure}
\centering
\centerline{\includegraphics[width=7cm]{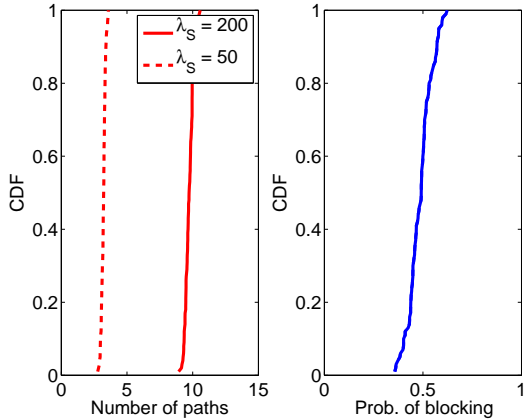}}
\vspace*{-.1in}
\caption{Left figure: CDF of the number of paths in user-RRH channels in coverage. Right figure: CDF of blocking probability in user-RRH pairs in coverage. }
\label{fig_numpaths}
\vspace*{-.1in}
\end{figure}

The left-hand-side figure in Fig. \ref{fig_numpaths} shows the distribution of the number of paths in the user-RRH channels in coverage for the two different values of $\lambda_S$. As the figure reveals, increasing $\lambda_S$ results in increased number of paths. The number of paths in our simulation environment plays a similar role to the ``number of clusters'' parameter in 3GPP's SCM specifications \cite{3DSCM}. In the SCM, this parameter is chosen between $12-20$ depending on the scenario considered in the  sub-6 GHz band. In this work, our parameter selection provides a lower number of paths, which might be  more appropriate for higher frequency bands.  The right-hand-side figure in Fig. \ref{fig_numpaths} shows the distribution of probability of blocking for LOS paths. In our setup, NLOS paths are not affected by blockers and hence the blocking distribution is not a function of $\lambda_S$.

\section{ Performance Evaluations}\label{perf-eval-sec}

In this section, we investigate the effect of sectors and directional training on area multiplexing gains and average uplink pilot transmit power spent for each packet. We provide various performance results of the proposed non-orthogonal pilot codes.
For evaluating performance, our simulations are based on geometric sectorization, i.e. radial sectorization with equal size wedges corresponding to an angle of $2\pi/S$ radiants per sector.

\subsection{Omni-directional Training}

\begin{figure}
\centering
\centerline{\includegraphics[width=7cm]{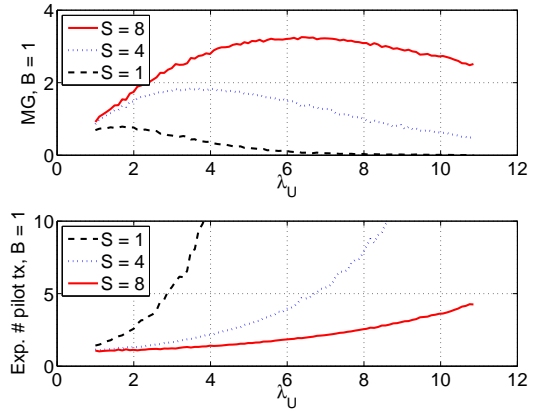}}
\vspace*{-.1in}
\caption{Top figure: Area multiplexing gains per pilot dimension vs. scheduling user intensity (omni UL pilots). Bottom figure: Expected number of UL pilot transmissions per delivered packet vs. scheduling user intensity (omni UL pilots).}
\label{fig_omni}
\vspace*{-.1in}
\end{figure}

In omni-directional training, users are assumed to emit uplink pilots isotropically. The upper figure in Fig. \ref{fig_omni} shows ${\rm MG(\lambda_U)}$ versus $\lambda_U$ for $B = 1$ for the cases where the number of sectors/RRH are $S = 1,\, 4$, and $8$.
Inspection of the figure reveals that, while the optimal scheduling intensity is $\lambda_U^* = 3$ for $S = 4$, it is almost $6$ for $S = 8$. The user-scheduling intensity optimized area multiplexing gains per pilot dimension are less than one with $S = 1$, but they are almost doubled with $S = 4$ and more than tripled with $S = 8$.

The increased multiplexing gains with non-orthogonal training come at the cost of more uplink pilot transmissions per delivered user streams. Indeed,  not all scheduled users (that transmit pilots) are  resolved and thus served in any given scheduling instance, because of pilot collisions. It is thus of interest to compare the relative average power spent for each packet transmission when comparing different schemes and parameter values. For orthogonal training, a user pilot transmission always results in a user being served provided the user is  not in outage. Thus, excluding users in outage, for orthogonal training, the expected number of pilot transmission per served packet is equal to $1$. 
With the schemes we consider, assuming a user UL pilot  (in a single slot)  that results in a user channel being resolved also enables serving a single packet,  we can capture the pilot of efficiency of a scheme via the expected number of pilot transmissions per each served packet.
The lower figure in Fig. \ref{fig_omni} shows the expected number of pilot transmissions per packet vs. $\lambda_U$ for different $S$ values. As the figure reveals, while the overall multiplexing gain is tripled with $S = 8$ (at $\lambda_U\approx 6)$, the transmit uplink power required per delivered packet is almost doubled compared to the orthogonal case. That is, on average two UL pilot transmission are required per user to deliver a packet.

\subsection{Directional training}

In this section we study how varying the user beam-width can affect the harvested area multiplexing gains. Indeed, using a directional beam at a user terminal makes its  user-channel sparser in terms of the number of sectors that are excited at the BS, thereby leaving more sectors available to resolve other users' channels.  In \cite{icc2017-zheda}, for a single cell scenario, a proper choice of the user beam width positively impacted both multiplexing gains and long-term user rates.

Directional beams with different width are obtained by processing at the user side similar to the sectorization at the RRH site. Each sector created at the user can be thought of as a beam candidate which the user randomly selects. The number of sectors created at the user antenna array, $B$ determines the width of the user beam. For radial sectorization, beam width is equal to $2\pi/B$. For the same outage probability, with directional training ($B>1$), the transmission power spent per $B$ pilot transmissions is equal to the uplink power spent for a single pilot transmission with omni-training $B = 1$.

\begin{figure}
\centering
\centerline{\includegraphics[width=7cm]{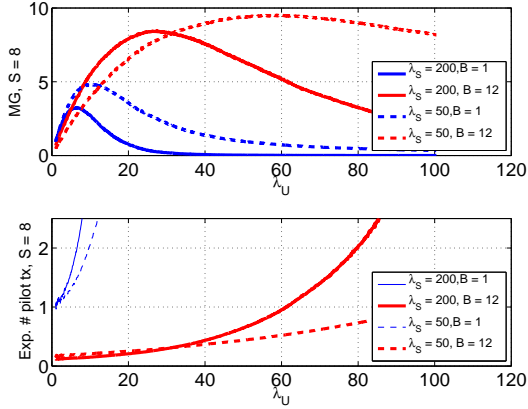}}
\vspace*{-.1in}
\caption{Top figure: Area multiplexing gains per pilot dimension vs. scheduling user intensity. Bottom figure: Expected required UL transmit power per delivered packet vs. scheduling user intensity. In both figures $S = 8$.}
\label{fig_directional}
\vspace*{-.1in}
\end{figure}

Fig. \ref{fig_directional} compares omni-directional training with directional training of $30^\circ$ beams, i.e., $B = 12$. This allows serving more than $8$ users per pilot dimension per unit area while using a quarter of the expected pilot transmission power per user with respect to the orthogonal case. Hence, using directional beams and non-orthogonal training is able to serve instantaneously many more users per unit area and at much less power cost to user devices.

\begin{figure}
\centering
\centerline{\includegraphics[width=7cm]{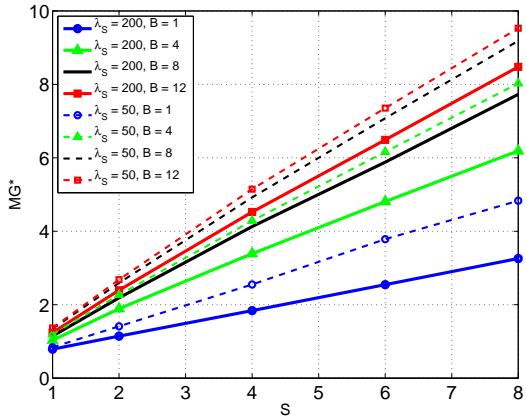}}
\caption{Scheduling user intensity optimized area multiplexing gains per pilot dimension as a function of the UL pilot beam-width  and the number of RRH sectors (geometric RRH sectorization). }
\label{mg_practical_antenna_designs}
\label{fig_allB_S}
\end{figure}

Fig. \ref{fig_allB_S} reports ${\rm MG^*}$ from (\ref{mq-opt}), i.e., the user-scheduling intensity optimized area multiplexing gains per pilot dimension, for various values of $S$ and $B$  and for two different scatterer-intensity scenarios. As the figure reveals,  uniformly higher area multiplexing gains are possible in the  lower scatterer-intensity environment.

\section{Area Multiplexing Gains with Practical Sectorization}\label{antenna-array}

In this section, we consider a PPP-based evaluation of  the area multiplexing gains that can be expected over  wireless networks of RRHs using practical antenna configurations and practical sectorization. Traditionally, sectors at  a macro cell are obtained by mounting a separate antenna array at the BS site for each sector. By using different array orientations in different sectors and by physically separating them  via panels, sector arrays can be made to ``cover'' different ``pie-type'' slices of the cell. For example, by using $P$ ULAs at a macro BS site, a macro cell (centered at the BS) can be divided into $P$ $360/P$-degree sectors. In the commonly used  case of $P = 3$, three 120-degree sectors are formed in the macro cell.
Such physical sectors were leveraged in \cite{icc2016-ozgun} to substantially reduce the number of RRH sites needed to achieve a given target multiplexing gain in a finite-area RRH deployment.

Besides traditional physical sectorization, virtual sectorization based on spatial processing at the BS antenna array is also possible \cite{sayeed-virtual},\cite{icc2017-zheda, Ansuman-JTSP-2014,adhikary-mmWave-jsac}.  Ref. \cite{icc2017-zheda} considers a single-cell scenario with a BS using a uniform linear array (ULA). DFT based processing is considered for virtual sectorization within the cell, according to which each sector is formed by spatially filtering the ULA signal through a contiguous set of sector-specific DFT beams.  Indeed, with  large ULAs,  DFT based processing  is the appropriate choice for sectorization, since, in the limit  of large ULAs, the user-channel covariance matrix becomes approximately circulant \cite{Ansuman-JTSP-2014,adhikary-mmWave-jsac}. In practice, power emission from an antenna array is not isotropic, and the power radiation patterns  highly depend on the antenna array shape and element spacing. In general,  although equal number of beams are assigned to each of the $V$ virtual sectors, the virtual-sector azimuth spread varies from sector to sector, in contrast to physical sectors.

In this section, we consider various combinations of physical and virtual sectorization options, which result in the same number of sectors $S$.  In particular, we consider combinations of $P$ physical sector sites per RRH, each with a ULA performing virtual sectorization in  $V$ virtual sectors, yielding a total of $S = VP$ sectors. The ULA-induced virtual sectorization scheme results in non-uniform sector sizes, which are the  result of the transformation between the normalized virtual angle, $\theta$ and the physical angle $\phi$. For a critically spaced ULA this transformation is given by $\theta = \pi\sin(\phi)+\pi$  \cite{sayeed-virtual},\cite{Bajwa-sparse}.

\begin{figure}
\centering
\centerline{\includegraphics[width=7cm]{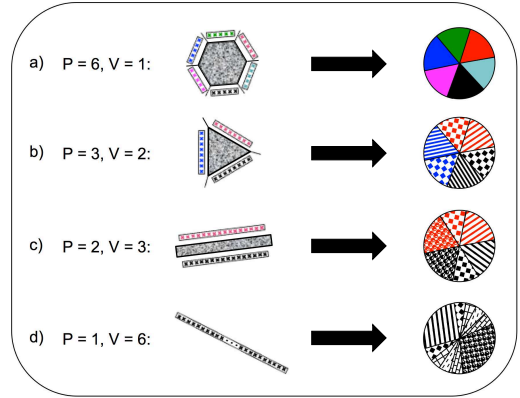}}
\caption{ Practical sectorization designs (left) and associated sectorization patterns (right).}
\label{fig_antenna_arrays}
\vspace*{-.1in}
\end{figure}Fig.~\ref{fig_antenna_arrays} depicts  several RRH designs exploiting various combinations of virtual and physical sectorization, all yielding $S=P\times V= 6$ sectors. All virtual sectors are created by critically spaced ULAs. For each ($P$, $V$) combination shown on the left-hand side of the figure, the corresponding azimuth support of  each of the  $6$ sectors is shown on the right-hand side. As the figure reveals, the $(P=6,\,V=1)$ and the $(P=3,\,V=2)$ designs yield the geometric 6-sector sectorization patterns. The $(P=3,\,V=2)$ design for instance relies on panels in the back of each ULA to restrict the paths hitting each ULA to come for a 120-degree azimuth spread.

\begin{figure}
\centering
\centerline{\includegraphics[width=7cm]{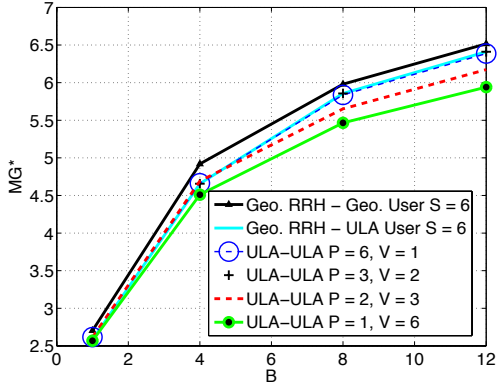}}
\caption{Scheduling user intensity optimized area multiplexing gains per pilot dimension as a function of (nominal) UL pilot beam-width (360/B degrees) for the practical sectorization designs in Fig.~\ref{fig_antenna_arrays}. }
\label{mg_practical_antenna_designs}
\vspace*{-.1in}
\end{figure}
Fig.~\ref{mg_practical_antenna_designs} shows the area multiplexing gains  of all these practical designs against those provided by geometric sectorization. With each practical design, a ULA is exploited at each user terminal for creating the $B$ directional beams.  Also shown for comparison is the performance with geometric sectorization at RRH and ULA at the user device.
As it can be seen from the figure, designs in a) and b) are identical in terms of sectorization to the geometric case (with ULA at the user device) and  provide the same area multiplexing gain performance. The $(P=2,\,V=3)$ and the $(P=1,\,V=6)$ designs yield a small loss in performance, revealing that the cost of virtual sectorization is inherently small.

\begin{figure}
\centering
\centerline{\includegraphics[width=7cm]{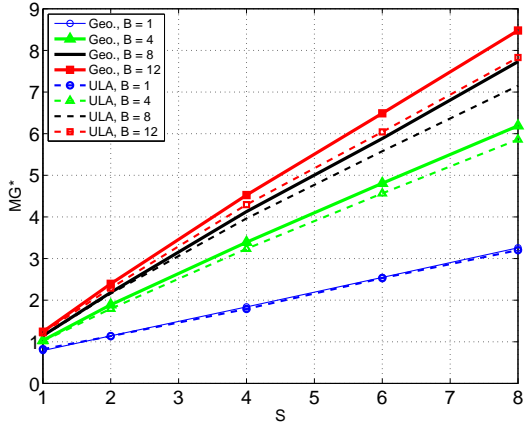}}
\caption{Scheduling user intensity optimized area multiplexing gains per pilot dimension as a function of number of RRH sectors and UL pilot training beam width: geometric (solid) and ULA based (dashed) sectorization. }
\label{fig_ULA_Geo}
\vspace*{-.1in}
\end{figure}
Finally, we compare the area multiplexing gains per pilot dimension that are provided by   a single critically spaced ULA with random orientation at each RRH, i.e., by employing  design option d) in Fig. ~\ref{fig_antenna_arrays} with $(P=1,\,V=S)$, against the ones provided by  the geometric $S$-sector RRH sectorization.
In the case of ULA-based RRH sectorization, users transmit directional pilots from a ULA able to create $B$ virtual-sector training beams, while in the geometric RRH sectorization case users employ the $B$ equal-width pilot beams considered in Sec.~\ref{perf-eval-sec}.
Fig.~\ref{fig_ULA_Geo} shows the area multiplexing gains per pilot dimension of the two schemes for various $(S,\,B)$ combinations.  As the figure shows, the geometric scheme, which results in uniform radial sectorization, achieves higher gains than the single-ULA  based  scheme. However, it can also be seen that the performance loss due to using  virtual sectorization is rather small for all $(S,\,B)$ combinations considered.


\bibliographystyle{IEEEtran}
\bibliography{IEEEabrv,refs}

\end{document}